\documentclass[sigconf]{acmart}

\usepackage{algorithmic}
\usepackage{graphicx}
\usepackage{textcomp}
\usepackage{xcolor}
\usepackage{multirow}
\usepackage{makecell}
\usepackage{ragged2e}

\usepackage{booktabs}
\usepackage{microtype}
\usepackage{tabularx}
\newcolumntype{R}{>{\raggedright\arraybackslash}X}
\usepackage[utf8]{inputenc}
\usepackage{placeins}
\usepackage[english]{babel}
\usepackage{csquotes}
\usepackage{xcolor}

\AtBeginDocument{%
  \providecommand\BibTeX{{%
    \normalfont B\kern-0.5em{\scshape i\kern-0.25em b}\kern-0.8em\TeX}}}

\copyrightyear{2020} 
\acmYear{2020} 
\setcopyright{acmlicensed}
\acmConference[ICSE-SEET'20]{Software Engineering Education and Training}{May 23--29, 2020}{Seoul, Republic of Korea}
\acmBooktitle{Software Engineering Education and Training (ICSE-SEET'20), May 23--29, 2020, Seoul, Republic of Korea}
\acmPrice{15.00}
\acmDOI{10.1145/3377814.3381708}
\acmISBN{978-1-4503-7124-7/20/05}

\begin{document}

\title{What prevents Finnish women from applying to software engineering roles? A preliminary analysis of survey data}

\author{Annika Wolff}
\affiliation{%
  \institution{LUT University}
  \city{Lappeenranta}
  \country{Finland}
}
\email{annika.wolff@lut.fi}

\author{Antti Knutas}
\affiliation{%
  \institution{LUT University}
  \city{Lappeenranta}
  \country{Finland}
}
\email{antti.knutas@lut.fi}

\author{Paula Savolainen}
\affiliation{%
  \institution{Turku University of Applied Sciences}
  \city{Turku}
  \country{Finland}
}
\email{paula.savolainen@turkuamk.fi}

\renewcommand{\shortauthors}{Wolff et al.}

\begin{abstract}
Finland is considered a country with a good track record in gender equality. Whilst statistics support the notion that Finland is performing well compared to many other countries in terms of workplace equality, there are still many areas for improvement. This paper focuses on the problems that some women face in obtaining software engineering roles. We report a preliminary analysis of survey data from 252 respondents. These are mainly women who have shown an interest in gaining programming roles by joining the Mimmit koodaa initiative, which aims to increase equality and diversity within the software industry. The survey sought to understand what early experiences may influence later career choices and feelings of efficacy and confidence needed to pursue technology-related careers. These initial findings reveal that women's feelings of computing self-efficacy and attitudes towards software engineering are shaped by early experiences. More negative experiences decrease the likelihood of working in software engineering roles in the future, despite expressing an interest in the field. 
\end{abstract}

\ccsdesc[500]{Social and professional topics~Computing occupations}
\ccsdesc[500]{Social and professional topics~Women}
\ccsdesc[300]{Social and professional topics~Computing profession}

\keywords{women in technology, careers, barriers, quantitative, survey, path modeling}

\maketitle

\section{Introduction}
The story presented through both the international and local Finnish media is that Finland has good gender equality. For example, based on EU Gender Equality Index 2019, Finland is overall one of the most gender-equal societies in Europe \cite{Eige}, 57.4 percent of persons having tertiary level degree are women, and employment rate of women aged 15 to 64 in 2017 is 68.5 percent (men 70.7 percent) \cite{Tila}. Whilst this is true in some respects, as evidenced in the next paragraph, broad statistics do not reflect the full picture and it is still the case that women face barriers to entry to some careers. If Finland is to live up to its reputation and improve this situation, it is important to first understand why this is the case and then to take practical steps to improve the situation in the future. 

One area where Finland could improve is in attracting and recruiting more women into technology-based roles, such as software engineering. In terms of women in science and technology, Finland compares unfavourably against many other European countries. A UNESCO science report from 2017 identifies that only 22 percent of graduates in engineering from Finnish universities are women \cite{baskaran2017unesco}. In terms of the overall job market, Eurostat statistics \cite{EuroS} from 2018 identify that only 28 percent of scientists and engineers in Finland are women, putting it almost in last place compared to other European countries (for example, in Lithuania the figure is 58 percent). The Finnish media itself reports on this trend, describing Finnish education choices as being strongly divided according to gender \cite{HSan}.

Often, if there are gender inequalities in a workplace it is tempting to look at the organisations who hire and retain staff, to focus on the hiring policies, the working culture and the support that is in place to prevent discrimination or hostile working environments.  However, what if women are not even applying to these positions in the first place  - even if they would be interested in them - due to their own perceptions and past experiences? What effect might such perceptions and experiences have had on their attitudes and feelings of efficacy, and therefore their confidence to apply for roles? Vehvilainen \cite{vehvilainen1999gender}, in exploring the reasons for disparity in the computing field in Finland reveals a history of male dominance in which all computer pioneers were men, coupled with the exclusion of women from visible public discourse and evidence of tokenism. While times are certainly changing, this history is still recent and it is not hard to imagine that it could still be affecting people’s attitudes today. These ideas about the root cause of gender inequality in technical professions being based on unconscious bias and gender stereotyping are backed up by a recent study, which also reveals that ideas of capability which have been created in childhood are significant when the young people think about their education choices \cite{Mika}.

This paper seeks to cast a light on what positive or negative experiences and attitudes may be influencing women’s willingness to actively apply for and enter software engineering roles in Finland and to understand when and where these problems usually begin. This paper reports on a survey designed to understand the experiences of 252 respondents who were mainly women who already work in, or are somehow already interested in applying for software engineering roles. 

\section{Background and hypotheses}

Our Research Question is \textit{'what experiences may directly or indirectly affect the willingness of Finnish women to apply for SE roles?’} We derived a number of hypotheses through exploring related work, which informed the design of a survey. In this section, we introduce related research in the field, the constructs from literature we explore, and the hypotheses that are related to those constructs.

The ‘leaky pipeline’ of STEM (Science, Technology, Engineering and Math) sees increasing gender disparity in STEM fields, including computing related subjects, across three stages of life \cite{dasgupta2014girls}. These are i) during childhood and the school years, when male stereotypes around STEM begin to emerge  ii) in early adulthood, when making decisions about courses in higher education  iii)  in early to mid adulthood, when making career choices and entering the workforce. Common explanations point to discrimination and stereotyping as a significant factor in the gender disparity both in educational and workplace settings.

\subsection{Early School Experiences}
Early studies looking for the causes of gender differences in attitudes to computing at school age have shown that at middle school age, boys are more likely to own computers, to use them both in and out of school and to hold gender stereotypical views about computers \cite{durndell1995gender}. More recent research \citep{ehrlinger2018gender, colley2003age} has shown that this trend continues - despite technology use increasing exponentially in intervening years - with women far less likely to relate to what they describe as the typical computer scientist than men, with the result that they are less interested in the field. Gender stereotypes can have a secondary effect, when gender stereotypes influence also the actions of teachers and caregivers \cite{cohoon2003must}. Outcomes include less opportunities to engage in computing clubs at school \cite{abbate2012recoding} and less encouragement from parents towards girls to participate in computing activities \citep{fox1982accelerated, nelson1991computer}.

\subsection{Higher Education Experiences}
Whether it is for these, or other reasons, a new problem starts to emerge. The attitudes of girls towards computing starts to vary. Attitude can be deconstructed to different aspects, including emotional affect, level of personal interest,  beliefs, general self-confidence and self-efficacy, which relates to a person’s belief in their own capability to perform a task \cite{cai2017gender}. Michie and Nelson \cite{michie2006barriers} posited that self-efficacy could play a crucial role in determining career choice and sought to understand whether this could explain under-representation of women in certain careers. Through empirical testing they discovered that women tended to report lower self-efficacy compared to men for occupations not typically associated with women (such as accountant, drafter, engineer, mathematician), even though more objective measures through test scores revealed no difference in ability.

Related studies focusing on aspects such as self-efficacy and confidence in computing, specifically, find that women are generally less likely to feel confident in using computers \citep{cai2017gender, young2000gender, burke2006barriers}. Crucially, for female students there is a correlation between their self-efficacy and their interest in a career in software programming \cite{aivaloglou2019early}. In high school, these attitudes become more entrenched, such that older females hold less positive attitudes towards computing than younger ones \cite{colley2003age}. These views are also shared by males, some of whom have a tendency to doubt the capabilities of women in IT \cite{burke2006barriers}.

\paragraph{Hypotheses H2a. - H2e.} Equal technology learning opportunities affect a) affective attitudes, b) behavioural attitudes, c) cognitive attitudes, d) computer self-efficacy, e) general self-efficacy.

\subsection{Working Life Experiences}
Despite these earlier experiences, many women still enter into technology-related professions and hiring policies, an area often put under the spotlight when diversity issues are considered, have been improving with efforts taken to minimise bias in a selection process. However, when considering issues of retention and progression within a company the picture is less clear. Typically masculine working cultures and lack of flexibility towards women’s often differing workplace needs can still be barriers \cite{servon2011progress}. Whilst Wentling and Thomas \cite{wentling2009workplace} highlight also the positive role that a workplace culture can play in a woman’s career development - for example, in companies that place an emphasis on skills training and collaborative working - they note that even in these cases these benefits may be offset by the more negative effects of male dominated and male focused working environments.

There are many initiatives that draw on research such as these outlined above to try and address issues at these three distinct phases of life. A full discussion of all the related issues and approaches to overcome them is beyond the scope of this paper, which offers only a glimpse of the discourse on this topic over the years. 

\paragraph{Hypotheses H1a. - H1e.} Negative experiences or encountering barriers in applying for software programming role affect a) affective attitudes, b) behavioural attitudes, c) cognitive attitudes, d) computer self-efficacy, and e) general self-efficacy.

\subsection{Beyond the Pipeline}
The ‘leaky pipeline’ metaphor and approach to identifying problems at transition points and focusing attention on these  has drawn some criticism \cite{vitores2016trouble}. Firstly, it must be noted that the gender gap clearly still persists, despite initiatives aimed at stopping the leak at these critical junctures. While there is evidence that it is closing in terms of who are users of computers and other technologies, it is still the case that fewer women than men are entering into professional computing roles \cite{faulkner2007gender}. Second, this model implicitly treats women, or other minorities, as ‘populations of untapped resources’ and thus diminishes their agency and ability to act towards their own interests \cite{vitores2016trouble}. Finally, the pipeline does not take into consideration that women’s career paths are far more likely to take a non-traditional and non-linear path than men’s \cite{caprile2010science, hyrynsalmi2019motivates}. Therefore, comparing boys with girls, or men with women at different stages of life may be misleading. A more fruitful approach may be to identify the root causes of why women who want to enter the profession - at any stage in life -  may feel more, or less, inclined to do so.

For these reasons, the research outlined in this paper focuses not on finding differences between men and women, or in trying to understand - at least at this stage - what prevents certain women from even considering technology related careers. Instead, we look at women who have already expressed an interest in technology-related careers and explore the various factors, such as attitude and self-efficacy, to understand what may affect their willingness to put themselves forward for roles. We try to identify when problems arise, but also to identify positive experiences and how they may have influenced career choice. The overall aim is to provide evidence to support initiatives that are trying to reduce inequality in computing fields.

\paragraph{Hypotheses H3a. - H3c.} a) Affective attitudes, b) behavioural attitudes, and c) cognitive attitudes have an effect on women's technology careers.

\paragraph{Hypotheses H4a. - H4b.} a) Computer self-efficacy and b) general self-efficacy have an effect on women's technology careers.

\paragraph{Hypotheses H5a.- H5e.} Negative experiences in applying or encountering barriers affect women's technology careers through the moderating factors in H1.

\paragraph{Hypotheses H6a.- H6e.} Equal technology learning opportunities affect women's technology careers through the moderating factors in H2.

\subsection{Research Model}
Our research model includes the theoretical constructs presented in this section and used in building the survey, and hypotheses about effects of one construct on another.

\begin{figure*}[ht]
    \centering
    \includegraphics[width=0.75\textwidth]{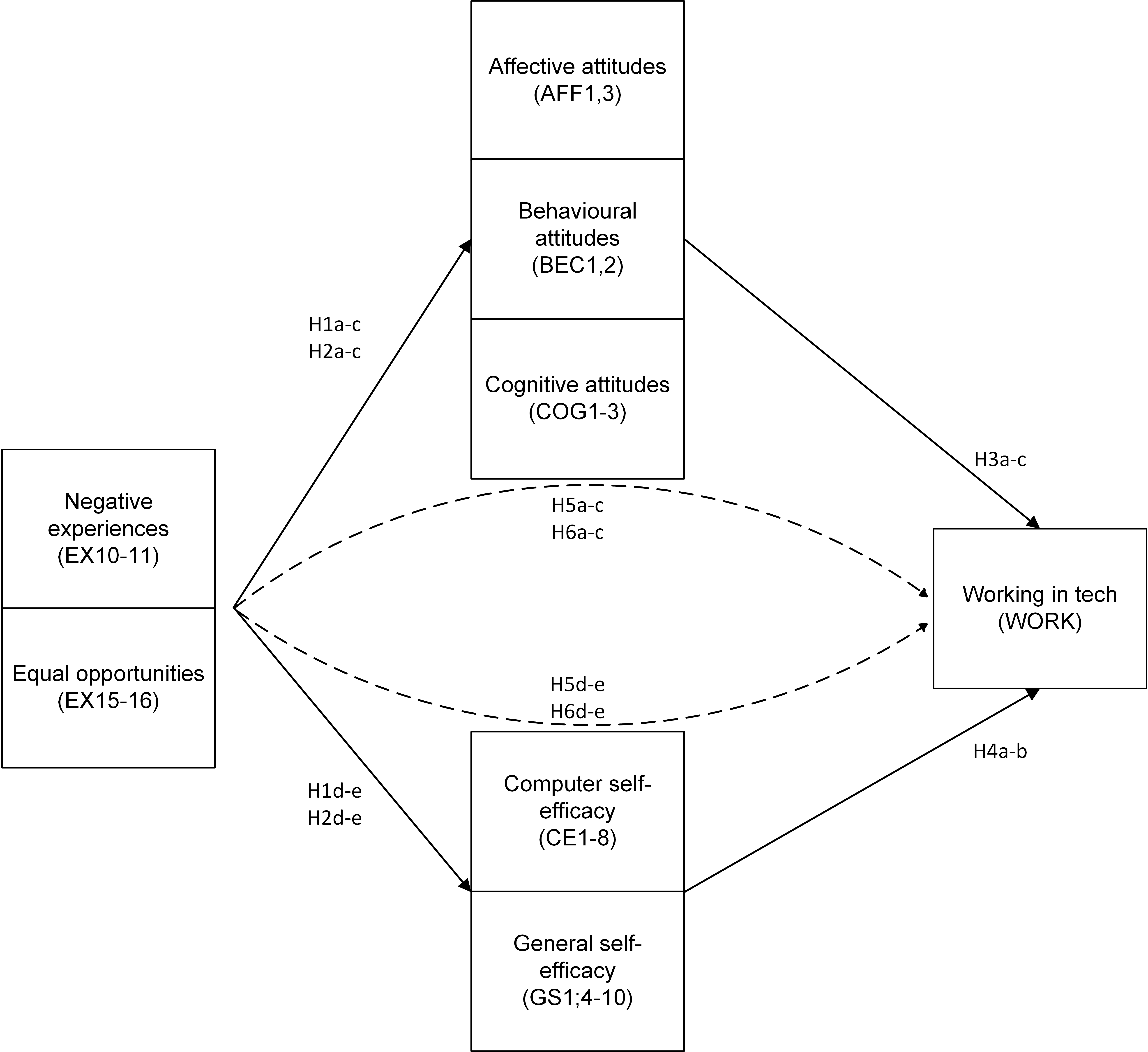}
    \caption{Research model with constructs and hypotheses}
    \label{fig:researchmodel}
\end{figure*}

The research model is visualized in Fig.~\ref{fig:researchmodel}.  Hypotheses about the effect of one construct on another are presented as arrows. For clarity, several parallel hypotheses are represented by a single arrow. Indirect effects are presented as dotted lines. Abbreviations inside the boxes refer to the specific survey items that were used to operationalize the construct and abbreviations next to the arrows refer to the specific hypotheses presented in this section.

\section{Methods}
Our research was conducted in cooperation with the Finnish Software and E-business Association, which has launched a Mimmit Koodaa program to ease the burning shortage of experts in the software industry by encouraging and training more women to work in the field.

To answer our research questions, the survey was created by adapting existing (validated) scales for measuring attitudinal components and self-efficacy, then by framing a series of both closed and open-ended questions for eliciting information on past experiences and by collecting demographic and other related information. Adaptation of the existing scales was for the purposes of i) tailoring existing attitude rating scales to the computing context ii) harmonising the Likert scales of different survey instruments to a 7 point scale. This step was considered necessary to improve the user experience and was possible as we were not directly replicating the studies and did not therefore need exactly the same Likert rating scale for comparison.

This resulted in a survey that had four distinct parts.
\begin{enumerate}
\item Part 1: attitudes towards women in software programming roles. This part consisted of 12 questions, aimed to measure cognitive, affective and behavioural components of attitude towards computing. For example, whether the respondents thought there were many opportunities in software programming roles in Finland and whether they were more likely to be taken by men (cognitive) whether they were anxious about computers or enjoyed learning new software skills (affective) and whether they had intentions to learn new skills or apply for software programming positions (behavioural). It was adapted from \cite{watson2016learning}.
\item Part 2: self-efficacy. This part was designed to measure both general \cite{schwarzer1995generalized} and computing \cite{laver2012measuring} self-efficacy.
\item Part 3: personal experience. This mixture of closed and open-ended questions were designed to discover respondents past or present experience in working in software engineering roles and uncovering past experiences both during school, higher education or working life.
\item Part 4: demographic questions. This section tailored standard demographic questions to a Finnish context.
\end{enumerate}

Similarly, our research model was based on three attitude constructs, two self-efficacy constructs, and two personal experience constructs.

The questionnaire consisted of 48 questions in the first 3 sections, plus the demographic questions. It is available as an online appendix\footnote{\RaggedRight Full questionnaire available at http://doi.org/10.5281/zenodo.3634590}, with abbreviations matching the constructs in Table \ref{tables/measurement}.

The questionnaire was designed in English and translated into Finnish. The respondent selected the language s/he wanted to use. It was delivered through the Webropol survey tool. Participation was voluntary and respondents were free to close the survey at any time and indicate that they did not wish their data to be used. 

The sections 1-4 were presented in a fixed order, while the questions within them were randomised to minimise order bias.

A pilot questionnaire was tested with 6 people working in the field and the survey was improved and clarified based on the feedback by the pilot group.

The questionnaire was open from 21.3.2019 until 30.4.2019. It was launched via Mimmit koodaa Facebook pages on 25.3.2019 and advertised in a newsletter sent by the association on 31.3.2019. 2039 persons liked the Mimmit koodaa Facebook pages and 2220 followed the pages on 1.4.2019. The Finnish newsletter was sent to over 3500 persons and English version to about 350 persons.

During this time, there were 254 respondents to the survey. Two people didn’t want to continue to participate. Therefore, the final number of respondents was 252 - 248 who identified themselves as women, 0 who identified themselves as men, 3 who identified as ‘other’ and one that did not answer the question. The majority of respondents (45 percent) were between 30-39. 22 percent were between 21-29,  21 percent were between 40-49 and 9 percent were between 50-59. The remainder were under 20 (1 percent) or over 60 (2 percent). The majority had completed some form of higher education (30 percent had a bachelors degree, or equivalent and 42 percent had a masters degree). Whilst 92 percent of respondents were currently living in a town, only 64 percent had attended high school in a town, demonstrating the trend towards urban living. 

\subsection{Analysis Methods}
We analyzed the research model with a path modelling approach. Path modelling allows specifying a visual research model with multiple theory-based constructs and defining paths between them that represent hypotheses of variable or construct relationships \cite{hair_jr_primer_2016}. We selected the Partial Least Squares Path Modeling (PLS-PM) statistical modelling method with non-parametric bootstrapping to cope with slight non-normality of the observed variables and low sample size (for a path modelling approach) \cite{hair_jr_primer_2016,henseler2014common,henseler2016using}. Furthermore, PLS-PM is an appropriate method to use because of the limited sample size  and the path model includes a formatively measured construct \cite{sarstedt2017partial}. We applied the SmartPLS 3 software package for data analysis \cite{ringle2015smartpls}.

\section{Analyzing the Path Model}
Our sample size was 252, which is enough for PLS-PM both according to the rule of thumb of having 10 times per largest number of paths from independent variable going into a dependent variable \cite{hair_jr_primer_2016} and using a power analysis guidelines adapted by Hair et al. \cite{hair_jr_primer_2016} from \cite{cohen1992power} at $5\%$ significance level.

For analysis purposes, the affective attitude construct was divided into interest (AFF1,3) and concern (AFF2,4) subconstructs, since the questions appeared to measure two different underlying phenomenona. Also, the variable COG3 was reversed, since the question was about the availability of opportunities, whereas COG1 and COG2 were about limitations to opportunities.

\subsection{Specifying and Validating the Measurement Model}\hspace*{\fill} \\

We assessed the validity of our measurement model by assessing the 1) measurement reliability, 2) model structure, and 3) discriminant validity following the guidelines by Hair and Henseler \cite{hair_jr_primer_2016,henseler2014common,henseler2016using}. Furthermore, we assessed nomological validity, according to which the relationships between the constructs in the path model, sufficiently well-known through prior research, should be strong and significant \cite{henseler_use_2009}. We used the outer loading relevance testing process by Hair \cite{hair_jr_primer_2016} to evaluate and exclude variables that had low outer loading to their respective constructs. Variables BEC3-4, COG4, and GS2-3 were excluded in the process.

In our model, we used both reflective constructs and one formative construct (WORK) \cite{hair_jr_primer_2016}. This causes some considerations, as not all validity measures apply to formative constructs \cite{hair_jr_primer_2016}. Assessing convergent and discriminant validity using criteria similar to those associated with reflective measurement models is not meaningful when formative indicators and their weights are involved \cite{chin1998partial,hair_jr_primer_2016}.

The measurement reliability was assessed with composite reliability (CR) and average variance extracted (AVE) indicating convergent validity. The measurement reliabilities are reported in Table~\ref{tables/measurement}. All of the convergent validity metrics were greater than the thresholds recommended in literature (CR $>$ 0.7; AVE $>$ 0.5) \cite{fornell1981evaluating,henseler2016using}. One construct was composed of one variable, which prevents the evaluation of its convergent validity. The fields where evaluation was not possible are marked with N/A. However, other quality metrics were high enough that we proceeded with analysis. Furthermore, single variable constructs are possible in PLS-PM, especially in cases where the variable is a concrete, observable item, even though their use decreases construct reliability \cite{hair_jr_primer_2016}.

\begin{table*}
\caption{Measurement reliabilities}\label{tables/measurement}
\centering
\begin{tabularx}{0.80\textwidth}{llllllll}
\toprule
 & Loading & t-value & p-value & Mean & SD & AVE & CR \\
\midrule
Affective attitudes & & & & & & 0.726 & 0.841 \\
AFF1. & 0.8437 & 5.978 & *** & 6.23 & 1.21 & & \\
AFF3. & 0.861 & 5.039 & *** & 6.06 & 1.17 & & \\
\midrule
Behavioural attitudes & & & & & & 0.812 & 0.896 \\
BEC1. & 0.857 & 24.084 & *** & 6.19 & 1.17 & & \\
BEC2. & 0.943 & 48.616 & *** & 5.10 & 1.87 & & \\
\midrule
Cognitive attitudes & & & & & & 0.533 & 0.764 \\
COG1. & 0.903 & 41.693 & *** & 5.10 & 1.75 & & \\
COG2. & 0.736 & 12.761 & *** & 5.80 & 1.58 & & \\
COG3. (rev) & 0.490 & 5.582 & *** & 5.96 & 1.15 & & \\
\midrule
Computer self-efficacy & & & & & & 0.659 & 0.939 \\
CE1. & 0.852 & 43.671 & *** & 4.65 & 1.92 & & \\
CE2. & 0.849 & 42.761 & *** & 3.81 & 2.19 & & \\
CE3. & 0.857 & 42.999 & *** & 4.22 & 2.14 & & \\
CE4. & 0.756 & 22.620 & *** & 4.23 & 2.30 & & \\
CE5. & 0.833 & 32.346 & *** & 3.69 & 2.18 & & \\
CE6. & 0.880 & 52.033 & *** & 4.40 & 1.99 & & \\
CE7. & 0.721 & 19.659 & *** & 4.12 & 2.07 & & \\
CE8. & 0.729 & 20.419 & *** & 3.90 & 1.90 & & \\
\midrule
General self-efficacy & & & & & & 0.697 & 0.948 \\
GS1. & 0.862 & 9.460 & *** & 5.47 & 1.27 & & \\
GS4. & 0.816 & 8.423 & *** & 5.50 & 1.25 & & \\
GS5. & 0.799 & 8.505 & *** & 5.67 & 1.17 & & \\
GS6. & 0.877 & 9.422 & *** & 5.95 & 1.04 & & \\
GS7. & 0.723 & 5.771 & *** & 5.19 & 1.37 & & \\
GS8. & 0.890 & 9.890 & *** & 5.61 & 1.21 & & \\
GS9. & 0.874 & 9.355 & *** & 5.76 & 1.13 & & \\
GS10. & 0.826 & 9.183 & *** & 5.55 & 1.25 & & \\
\midrule
Negative experiences & & & & & & 0.619 & 0.764 \\
EX10NEG. & 0.741 & 4.959 & *** & 1.21 & 0.41 & & \\
EX11NEG. & 0.831 & 6.914 & *** & 1.55 & 0.50 & & \\
\midrule
Equal opportunities & & & & & & 0.734 & 0.847 \\
EX15EQOP. & 0.854 & 26.905 & *** & 1.38 & 0.49 & & \\
EX16EQOP. & 0.860 & 27.671 & *** & 1.40 & 0.49 & & \\
\midrule
Working in a tech. career & & & & & & N/A & N/A \\
WORK & 1 & N/A & N/A & 1.26 & 0.44 & & \\
\bottomrule
\multicolumn{8}{X}{\scriptsize *) Statistically significant at p$<$0.05, **) Statistically significant at p$<$0.01, ***) Statistically significant at p$<$0.001}
\end{tabularx}
\end{table*}

We analyzed the structure of the measurement model, where applicable, by significance and weight of factor loadings, and for cross-loadings between the latent constructs. Loadings in the outer model (measurement model) were significant and varying from acceptable .741 to good .943, indicating valid model structure, with one exception. The loading for variable COG3 was lower than usually accepted (0.490). However, since the construct is from a pre-validated survey, was significant, and other construct quality indicators were satisfactory, the variable was retained.

We assessed discriminant validity of the measurement model, firstly, by the square root of AVE (i.e. Fornel-Larcker criterion \cite{fornell1981evaluating}), where all of the AVE square roots should be greater than the squared latent variable correlations. Secondly, we verified that all item loadings exceed cross-loadings \cite{henseler2016using}. Affective attitude subconstruct (AFF2,4) had to be excluded because its high heterotrait-monotrait ratio of correlations (HTMT) with the cognitive attitude construct, indicated by a a high HTMT value. After excluding the (AFF2,4) subconstruct, we verified that other constructs remained below recommended values (HTMT $<$ 1) \cite{henseler2015new}.

\subsection{Validating the structural model}\hspace*{\fill} \\
We used a bootstrapping method to evaluate the coefficients for their significance \cite{davison1997bootstrap}, which is the recommended method for PLS-PM \cite{hair_jr_primer_2016}. The bootstrap sample size was $n=252$, and resampling was performed 5000 times, which are parameters following best practises in literature and are sufficient to evaluate the model \cite{hair_jr_primer_2016,henseler2016using}. We tested and validated the quality of the structural model representing our hypotheses by evaluating 1) collinearity issues and overall fit, 2) explanatory power, and 3) path significances.

We assessed the collinearity and the model fit in order to validate the structural model and identify misspecification problems. VIF (variance inflation factor) of the latent constructs did not indicate collinearity issues with values clearly between recommended values (0.2 $<$ VIF $<$ 5) \cite{hair_jr_primer_2016}. Because of the hypothesis testing objective of the paper, we further assessed the overall fit of the structural model to the data in order to analyze whether the model was specified correctly. For that purpose, we used the standardized root mean square residual quality metric (SRMR $<$ .08) \cite{henseler2016using} to evaluate estimation error and misspecification of the model \cite{hair_jr_primer_2016}. In our case, SRMR $=$ .077 indicates a satisfactory model fit..

We present the path significances and hypothesis testing results in the next section.

\section{Findings}

In this section, we review the hypotheses testing results and then review a selection of open-ended answers to give additional context to the quantitative findings.

\subsection{Hypothesis Testing Results}
In path model -based hypothesis testing, each path between a construct is a hypothesis of one construct having an effect on another. The summary of the effects and therefore of hypothesis testing outcomes is presented in Table~\ref{tables/hypotheses}. Most important feature to review from the table is whether the theorized path is significant - meaning that the dataset had evidence to support to existance of the theorized effect. The second most important feature to review is the effect size $f^2$. We used guidelines by \cite{henseler_use_2009} for interpreting $f^2$: 0.02, 0.15, and 0.35, representing small, medium, and large effect.

\begin{table*}
\caption{Effects and hypothesis support}\label{tables/hypotheses}
\centering
\setlength{\tabcolsep}{3pt}
\begin{tabularx}{0.90\textwidth}{lllllll}
\toprule
Hypothesis & Path & Effect ($\beta$) & Effect size ($f^2$) & t-value & p-value & Supported \\
\midrule
H1a. & Negative experiences $\rightarrow$ Affective att. & 0.141 & 0.020 & 1.980 & * & Yes \\
H1b. & Negative experiences $\rightarrow$ Behavioural att. & 0.151 & 0.023 & 2.041 & * & Yes \\
H1c. & Negative experiences $\rightarrow$ Cognitive att. & 0.347 & 0.164 & 4.610 & *** & Yes \\
H1d. & Negative experiences $\rightarrow$ Computer S-E & 0.220 & 0.060 & 3.301 & ** & Yes \\
H1e. & Negative experiences $\rightarrow$ General S-E & 0.058 & 0.003 & 0.652 & n & No \\
\midrule
H2a. & Equal opportunities $\rightarrow$ Affective att. & 0.057 & 0.003 & 0.777 & n & No \\
H2b. & Equal opportunities $\rightarrow$ Behavioural att. & 0.074 & 0.006 & 1.120 & n & No \\
H2c. & Equal opportunities $\rightarrow$ Cognitive att. & -0.386 & 0.203 & 7.428 & *** & Yes \\
H2d. & Equal opportunities $\rightarrow$ Computer S-E & 0.376 & 0.175 & 6.875 & *** & Yes \\
H2e. & Equal opportunities $\rightarrow$ General S-E & 0.046 & 0.002 & 0.641 & n & No \\
\midrule
H3a. & Affective att. $\rightarrow$ Work & -0.188 & 0.030 & 2.376 & * & Yes \\
H3b. & Behavioural att. $\rightarrow$ Work & 0.170 & 0.022 & 2.380 & * & Yes \\
H3c. & Cognitive att. $\rightarrow$ Work & -0.164 & 0.026 & 2.377 & * & Yes \\
\midrule
H4a. & Computer S-E $\rightarrow$ Work & 0.382 & 0.115 & 5.769 & *** & Yes \\
H4b. & General S-E $\rightarrow$ Work & 0.064 & 0.005 & 1.125 & n & No \\
\midrule
H5a. & Negative experiences $\rightarrow$ Affective att. $\rightarrow$ Work & -0.026 & N/A & 1.525 & n & No \\
H5b. & Negative experiences $\rightarrow$ Behavioural att. $\rightarrow$ Work & 0.026 & N/A & 1.540 & n & No \\
H5c. & Negative experiences $\rightarrow$ Cognitive att. $\rightarrow$ Work & -0.057 & N/A & 2.171 & * & Yes \\
H5d. & Negative experiences $\rightarrow$ Computer S-E $\rightarrow$ Work & 0.084 & N/A & 2.895 & *** & Yes \\
H5e. & Negative experiences $\rightarrow$ General S-E $\rightarrow$ Work & 0.004 & N/A & 0.462 & n & No \\
\midrule
H6a. & Equal opportunities $\rightarrow$ Affective att. $\rightarrow$ Work & -0.011 & N/A & 0.463 & n & No \\
H6b. & Equal opportunities $\rightarrow$ Behavioural att. $\rightarrow$ Work & 0.013 & N/A & 1.024 & n & No \\
H6c. & Equal opportunities $\rightarrow$ Cognitive att. $\rightarrow$ Work & 0.063 & N/A & 2.150 & * & Yes \\
H6d. & Equal opportunities $\rightarrow$ Computer S-E $\rightarrow$ Work & 0.144 & N/A & 4.357 & *** & Yes \\
H6e. & Equal opportunities $\rightarrow$ General S-E $\rightarrow$ Work & 0.003 & N/A & 0.453 & n & No \\
\bottomrule
\multicolumn{7}{X}{\scriptsize n) not significant, *) Statistically significant at p$<$0.05, **) Statistically significant at p$<$0.01, ***) Statistically significant at p$<$0.001}
\end{tabularx}
\end{table*}

The hypotheses on negative experiences affecting all attitudes and computer self-efficacy were supported. Negative experiences having an effect on affective and cognitive attitudes was expected, since these attitude questions express negative sentiments. For example, affective questions include a question such as ``I feel anxious about using computers and other technologies'' and cognitive questions include ``it is easier for men to get software programming roles than women in Finland''. What was unexpected that the negative experiences also had a connection to computer self-efficacy and behavioural attitudes, which includes a question such as ``I intend to learn new software programming skills''.

The hypotheses on equal opportunities affecting attitudes were supported only on cognitive attitudes. Equal opportunities reducing cognitive attitudes was expected, since learning opportunities was theorized to be a positive experience that reduces a negative attitude. Equal opportunities also had a positive effect on computer self-efficacy.

All attitudes had a statistically significant effect with a small effect size on the respondent currently working in technology industry. Positive attitude (behavioural) had a positive effect and negative attitudes (behavioural and cognitive) had a negative effect.

In order to evaluate mediation effects, we follow a process by Zhao et al. \cite{zhao2010reconsidering}, which is also proposed by Hair et al. \cite{hair_jr_primer_2016} for use with PLS-PM. We used this process to evaluate the hypotheses that attitudes and self-efficacy mediate the effect between 1) equal learning opportunities and working in tech, and 2) negative experiences and working in tech. First, we establish that a minimum of one of the mediating effects is significant (H5c-d, H6c-d). Second, we establish that the direct effects are non-significant between 1) equal learning opportunities and work ($\beta = -0.040$, $p = 0.577$), and 2) negative experiences and work ($\beta = 0.101$, $p = 0.148$). Therefore the effect is fully mediated, meaning that the effect occurs only indirectly through the mediating construct. We can conclude that the hypotheses on equal learning opportunities and negative experiences having an effect on working in a technology career through cognitive attitudes and computer self-efficacy are supported.

As expected, equal learning opportunities had a positive effect effect on work through the mediating constructs. Also, it was expected that the negative experiences had a negative effect on working in technology careers through cognitive attitudes. What was unexpected that negative experiences had again a positive effect through computer self-efficacy.

The explanatory power $R^2$ for the latent variables in the path model varies from "General self-efficacy" $=$ 0.006 to "Work" $=$ 0.245, which range from non-existant to moderate according to guidelines interpreted by Henseler et al. \cite{henseler_use_2009} from \cite{chin1998partial}, indicating that not all constructs are relevant in the inner path model structure. The construct with moderate explanatory power were cognitive attitudes ($R^2 = 0.267$) and tendency to work ($R^2 = 0.245$). Computer self-efficacy had low explanatory power ($R^2 = 0.191$). Constructs with non-meaningful explanatory power were affective attitudes ($R^2 = 0.023$), behavioural attitudes ($R^2 = 0.028$), and general self-efficacy ($R^2 = 0.006$).

$R^2$ in the PLS-PM approach indicates how well the predicted constructs (endogenous constructs) are explained by the connected constructs (exogenous constructs). For example, how well the attitudes or self-efficacy are explained by equal technology learning opportunities or negative experiences. The lack of explanatory power in some constructs is a finding in itself, since it indicates that general self-efficacy and affective or behavioural attitudes are not as relevant constructs in future studies, compared to cognitive attitudes and computer self-efficacy. Also, the moderate to low explanatory power indicates that future studies should evaluate other exogenous constructs that would explain general self-efficacy and the two attitude types better. The model was not revised in this study, since adjusting hypotheses after statistical analysis findings\footnote{Often known as HARKing (hypothesizing after results are known) \cite{kerr1998harking}} is against best practises.

\subsection{Exploring Qualitative Findings}
Our research question concerned ‘what experiences may directly or indirectly affect the willingness of Finnish women to apply for SE roles?' The survey has provided rich qualitative data from open-ended questions that will be fully analysed and integrated with the quantitative data in future work. As a first step, and in order to provide some additional context to our quantitative data analysis in this paper, we have taken a first pass over the data and identified a selection of the  most characteristics quotes for each construct that was found to have a statistically significant impact on the tendency to work in technology careers.

\paragraph{Negative experiences}\hspace*{\fill}
\begin{displayquote}
    ``Dismissing me because of my gender; some actual eye-rolling and verbal belittling about my degree.''
\end{displayquote}

\begin{displayquote}
    ``As a young developer it became clear to me that you can make it only on the guys' terms. I wouldn't stay there being ridiculed at and I moved elsewhere.''
\end{displayquote}

\paragraph{Equal learning opportunities}\hspace*{\fill}
\begin{displayquote}
    ``They were really supporting at high school and I felt that I was encouraged for studying IT. In mathematics there were no similar support and I felt that's where I would have needed encouragement.''
\end{displayquote}

\begin{displayquote}
    ``We girls were not encouraged to go for those.''
\end{displayquote}

\begin{displayquote}
    ``Everyone had the same optional courses.''
\end{displayquote}

\paragraph{Attitudes}\hspace*{\fill}
\begin{displayquote}
    ``The problem really is that women even aren't seeking technology careers.''
\end{displayquote}

\begin{displayquote}
    ``I believe that walls and glass ceilings could be broken by showing that programming does not have to be the your entire life. It can be just like any skill.''
\end{displayquote}

\begin{displayquote}
    ``I feel all the time that I have to constantly prove that I really want this career and that I can really perform the tasks. I face harder critique because there is already a preconception that I couldn't.''
\end{displayquote}

\paragraph{Self-efficacy}\hspace*{\fill}
\begin{displayquote}
    ``I studied technology-related subjects, because they were interesting and no one prevented or criticized my choices. I did not need or want any special support.''
\end{displayquote}

\begin{displayquote}
    ``I always liked mathematics, physics, and chemistry lessons and I was good at them. Programming could have fitted in with them.''
\end{displayquote}

\section{Discussion}
In this paper, we present our preliminary analysis of a survey designed to discover how early experiences affect Finnish women's confidence to apply for software engineering roles. Our sample are a group of mainly females who have already expressed interest in this field and of whom some have already obtained programming roles. We focus this first analysis on understanding the differences between those who have taken the step and obtained roles and those who have not yet done so.

We found that negative early experiences and the perceived lack of equal opportunities in school did have an impact on the likelihood that someone would be working in a software engineering role in the future. In our path model, this was fully mediated by the impact such experiences would have on womens' feelings of computing self-efficacy and their attitudes towards the field. This effect was particularly strong for the cognitive attitudes and perceptions of the field as being male dominated. It is also important to note that general self-efficacy did \textit{not} have an effect, which suggests that these same women felt confidence in themselves and their ability to solve every day problems but that it was specifically their computing self-efficacy that was affected by earlier experience and that influenced their likelihood to find work in software engineering in the future.

These findings support many of the earlier studies \citep{cai2017gender, young2000gender, burke2006barriers} and show that even in the Finnish context, where statistics show good overall gender equality and equal access to roles, the gendering of specific roles still exists. This leads to a situation where young people grow up with the idea that they can do any job, but if they move outside their gendered roles they soon discover this is not the case. In fact, the reality seems to be that the more equal a society is, the less women graduate in STEM fields \cite{YLE} - and as our survey seems to show, this is not due to having a freer choice but rather down to negative experiences and gender stereotyping. 

Our \textit{main novel contribution} is for practise: We have identified some of the barriers that women face in Finnish software engineering careers and this can inform future research and policies. Our \textit{second novel contribution} is to the scientific body of knowledge in providing more evidence about i) how the barriers affect career choice through moderating constructs and ii) comparing these constructs though path modeling. Many studies, such as \cite{burke2006barriers}, investigate the direct effects of these barriers. In our study, we found out that negative experiences and equal learning opportunities had no direct effect on career choice and instead their impact was through their impact on attitudes and computing self-efficacy. This indicates that quantitative analysis should proceed to explore more complex models, such as the path models proposed by Ahuja \cite{ahuja2002women} in a literature review and explored by Armstrong et al. \cite{armstrong2014barriers} in a qualitative study. Currently many of these models have been proposed in literature, but not yet empirically tested through quantitative methods. Pursuing this approach would allow identifying barriers that have indirect or more granular effects.

Our findings support the idea that \textit{improving the situation for women in software engineering requires broad strategies that seek to change perceptions of society as a whole about who can and cannot participate in tech}. Indeed, Cheryan et al. \cite{cheryan2013stereotypical} have shown that popular media representations of women in computer science fields have the power to overcome stereotypes and can increase interest of women in the field. But beyond this, changing stereotypes has the possibility to reduce many of the problems identified within our work. In early school, it could ensure that not only are the equal opportunities for girls and boys at school, but that the girls feel comfortable in making choices according to their own interests rather than what they perceive as the expectation of their peers, parents, teachers and society as a whole. This is not to say that girls should be forced to participate in such activities in equal numbers as the boys, only that their participation should be seen and treated as being completely normal. This could lead to improved computing self-efficacy and more confidence to enter higher education or apply for technical roles. Further, changing stereotypes could lead to a reduction in gendered advertising for roles and improve workplace cultures \cite{patitsas2014historical}. 

As identified in our work, though, the situation could be set to improve. In looking to the media to understand the current discourse and in working with an initiative such as Mimmit koodaa we can see that awareness of the issues are leading to people to ask for change.

\section{Limitations and Future Work}
This paper has focused on the experiences of mostly just women, who have joined an initiative that has the direct focus of increasing the number of women working in technical roles within software companies. It therefore is likely to exclude the opinions of people who do not see it necessarily as a gendered issue, or are happy in their career choice already and do not feel the need to join an initiative focused on change. We also do not capture opinions from those who are not interested in software engineering in general, even if they may have been under different circumstances. However, there has been plenty of prior work trying to look at different stages of life what might be happening in these cases to disenfranchise people from the profession so early. As we only surveyed women within Finland, we cannot speak to what might be happening in other countries with different cultures and we do not, at this stage, try to understand how men see this issue.

What is  notable in our analysis, is that general self-efficacy had no effect size or meaningful explanatory power in our research model. This means that within our dataset it had no impact on a person working in a technology career and also that it was not explained by negative experiences or perceived learning opportunities. This indicates that computer and general self-efficacy are grounded in different experiences and future research should consider increased granularity. A similar phenomenon was observed to a lesser extent between different attitude types. Also, the medium to low explanatory power of mediating constructs indicates that more causes need to be explored beyond negative experiences and perceived learning opportunities.

Future work will explore the existing data in more detail and also we will look at expanding the survey to other contexts.

\section{Conclusion}
There are many reasons why women may find it harder than men to enter technical professions. In software engineering, prior research has identified a range of issues that begin in early school days and continue into working life. These are set against a backdrop of societal expectations and gender stereotyping. This work has explored these issues from the Finnish perspective. We look at the general view of Finland as a country with good gender equality. However, in reality many professions are heavily gendered, even more so than in countries with less gender equality overall. By surveying 252 women (including 3 non-binary persons) who either are working or are interested to work in software programming roles and asking about their experiences and opinions, we find that women do often feel discriminated against and that equal opportunities in school is variable. Through statistical analysis of the data, we find support for other researchers work and show that negative experiences can affect women’s feelings of self-efficacy and their attitudes towards computing as a career. We suggest that while there are a range of activities that will help the situation for women in Finland, of primary concern is tackling the stereotyping of computing as being a male field as this would have most impact across many of the issues identified at different stages of school life and working career. We look to the media as one possible route for change.

\FloatBarrier

\bibliographystyle{ACM-Reference-Format}
\bibliography{sample-base,anttis_references}

\end{document}